\documentclass[amsmath,amssymb,amsbsy,reprint,prb,preprintnumbers,showpacs,superscriptaddress]{revtex4-2}
\usepackage{graphicx,color}
\usepackage{dcolumn}
\usepackage{bm}
\usepackage{braket}
\usepackage{mathtools}
\usepackage{ulem}
\usepackage[breaklinks,colorlinks=true,linkcolor=blue,urlcolor=blue,citecolor=blue]{hyperref}
\usepackage{times}
\usepackage{physics}
\usepackage{latexsym}
\usepackage{amsmath, amssymb}
\usepackage{mathtools}
\usepackage{multirow}

\newcommand{\be}{\begin{eqnarray}}
\newcommand{\ee}{\end{eqnarray}}

\begin{document}

\title{Phase transition and evidence of fast-scrambling phase in measurement-only quantum circuit}
\date{\today}
\author{Yoshihito Kuno} 
\thanks{These authors equally contributed}
\affiliation{Graduate School of Engineering Science, Akita University, Akita 010-8502, Japan}
\author{Takahiro Orito}
\thanks{These authors equally contributed}
\affiliation{Insitute for Solid State Physics, The University of Tokyo, Kashiwa, Chiba, 277-8581, Japan}
\author{Ikuo Ichinose} 
\affiliation{Department of Applied Physics, Nagoya Institute of Technology, Nagoya, 466-8555, Japan}


\begin{abstract}
Information scrambling is nowadays one of the most important topics in various fields of research. 
Measurement-only circuit (MoC) exhibits specific information scrambling dynamics, 
depending on the types of projective measurements and their mutual anti-commutativity. 
The spatial range of the projective measurements in MoCs gives significant influences on circuit dynamics. 
In this work, we introduce and study long-range MoCs, which exhibit an interesting behavior in their dynamics.
In particular, the long-range measurements can induce volume-law phases in MoCs without unitary time evolution, 
which come from anti-commutative frustration of measurements specific to the long-range MoCs. 
This phenomenon occurs even in MoCs composed of solely two-body measurements, and it accompanies an entanglement phase transition. 
Crucially, our numerics find evidences that MoCs can be a fast scrambler. 
Interplay of high anti-commutativity among measurements and their long-range properties generates fast entanglement 
growth in the whole system beyond linear-light-cone spreading. 
\end{abstract}


\maketitle
\section{Introduction}
Measurement to quantum system leads to nontrivial dynamics and produces exotic phases of matter. 
One of the most interesting phenomena induced by measurements is entanglement phase transition in hybrid random unitary circuits, 
which is recently studied extensively \cite{Li2018,Skinner2019,Li2019,Vasseur2019,Chan2019,Szyniszewski2019,Choi2020,Bao2020,Gullans2020,Jian2020,Zabalo2020,Sang2021,Sang2021_v2,Nahum2021,Sharma2022,Fisher2022_rev,Block2022,Richter2023,Sierant2023}. 
Generally, projective measurements suppresses spread of entanglement. Competition between unitary time evolution and projective measurement induces a phase transition. 
This phenomenon emerges in various hybrid circuits including time-evolution operator of many-body Hamiltonian as unitary \cite{Fuji2020,Goto2020,Tang2020,Lunt2020,Turkeshi2021,Kells2022,Fleckenstein2022,KOH2022}. 
The properties of Hamiltonian such as integrability, long-range property of couplings etc., also induce rich dynamical phenomena due to their interplay with measurements. 

At present, quantum information dynamics is one of the most attractive topics in broad field of physics. 
How initial local quantum information in a quantum many-body system spreads and how fast it propagates into the entire system are problems under active study. 
The spread of quantum information is called scrambling. 
Initially, black hole was a target physical system \cite{Hayden2007}. 
Now understanding scrambling for various quantum systems is on-going issue \cite{Hosur2016,Xu2022}.

Applying quantum measurements to many-body systems leads to interesting dynamics. 
This originates from non-commuting nature between certain measurement operators in quantum mechanics.
When many-body quantum system evolves under continuous measurements, the system exhibits specific dynamics and generates strongly-correlated states, which are not induced by simple unitary dynamics. 
In quantum circuit, especially Clifford circuit, which is efficiently described by stabilizer formalism \cite{Gottesman1997}, projective measurement induces non-trivial dynamics. 
Recently it has been reported that, without time-evolution unitary operators, measurement-only quantum circuit (MoC) \cite{Lang2020,Ippoliti2021} displays striking phenomena. 
Interplay of different kinds of measurements, some of which are not commute with each other, 
exhibits counter-intuitive phenomena, which include novel entanglement phase transitions, emergence of non-trivial states such as measurement-only thermal state and critical phase, etc \cite{Ippoliti2021,Zhu2023}. 
Also, under a suitable choice of the measurement type, the MoC generates interesting topological phases of matter such as symmetry protected topological (SPT) state \cite{Lavasani2021,Klocke2022,KI2023} and topological order \cite{Lavasani2021_2,Negari2023}. 

As an interesting aspect of MoC, if a counterpart Hamiltonian could be constructed by using each measurement operator in a MoC, 
the steady states in the MoC would be very close to the ground state of the counterpart Hamiltonian system. 
Some numerical examples supporting the above conjecture have been reported for a projective transverse-field Ising model \cite{Lang2020} and some SPT Hamiltonian 
systems \cite{Lavasani2021,Klocke2022,KI2023}. 

Here, we pose an interesting issue, that is, for a MoC exhibiting a volume-law thermal phase, whether additional measurements can induce fast
spreading of entanglement or not.
In particular, fast scrambling, which was first studied in the context of black hole physics \cite{Sekino2008,Maldacena2016}, is of our interest. 

Our motivation of studying long-range measurements comes from the recent theoretical 
and numerical investigations on dynamics of system with (spatially) long-range unitary. 
Among them, Refs.~\cite{Yao2016,Belyansky2020,Bentsen2020,Li2020,Li2022} showed that fast scrambling occurs in spin-chains with long-range interactions, and 
Refs.~\cite{Sharma2022,Block2022,Hashizume2022,Hashizume2022_2,Kuriyattil2023,Richter2023} studied random Clifford circuit with measurement to find that
long-range unitary gives significant effects to entanglement dynamics. 
In addition, the recent studies of MoC showed that the counterpart Hamiltonian can be constructed for certain MoCs and it 
gives a significant insight about steady states of the corresponding MoC~\cite{Lang2020,Lavasani2021,Klocke2022,Sriram2022,KI2023}. 
From these previous studies, it is natural to ask how long-range measurements affect to the circuit dynamics, especially to entanglement and quantum correlations. 
Although effects of multi-body measurements has been investigated in Ref.~\cite{Ippoliti2021}, spatial long-range effects has not been clarified yet. 
Since long-range unitary (or all-to-all interactions~\cite{Lucas2019,Belyansky2020}) induces significant scrambling effects, 
we also expect that some frustrated long-range multi-body measurements 
can produce fast scrambling phenomena; \textit{possibility of fast scrambling of entanglement in MoCs}.
In this paper, we investigate this problem numerically. 
We employ efficient numerical simulation based on the Gottesman-Knill theorem \cite{Gottesman1997,Aaronson2004,Nielsen_Chuang}, 
which can be tractable to large system size, suitable platform to investigate spatial structure of stabilizer states. 

Throughout this work, we investigate effects of long-range measurements in MoC. 
To clarify the issues explained in the previous paragraph, we systematically study two and three-body measurement models 
of one-dimensional spin chains (one-dimensional qubits) by investigating various physical quantities suitable for stabilizer formalism. 
Our numerics find an interesting phase transition for these MoCs, 
coming from the interplay between high frustration nature among measurement operators and their
long-range spatial support.
We further find evidences of fast scrambling phenomena in the MoCs. 
We show numerical results to support the above findings throughout this work. 

Here, we would like to emphasize that the target circuits consist of the long-range measurements only, 
and it is an open problem how states behaves under `time evolution' in these circuits.
Furthermore, we expect that this type of circuits supply a benchmark test for the error tolerance of 
near-term noisy computers as the projective measurement is tractable and stable in most of cases.

The rest of this paper is organized as follows. 
In Sec.~II, we introduce two kinds of target MoC models, i.e., long-range two-body measurement model (LR2BM) and long-range three-body measurement model (LR3BM). 
In Sec.~III, physical quantities of our interest are explained. 
We display results of the numerical calculation and discuss their physical indications. 
We focus on the identification of steady states, and observe that both the models exhibit a volume-law phase. 
In particular, we find a phase transition to the volume-law phase, details of which are examined including its criticality. 
Then, in Sec.~IV, we clarify properties of scrambling for both the models. 
To this end, three approaches are employed. 
(I) We observe entanglement dynamics in an intermediate-time period.
(II) We evaluate a wave front dynamics by using entanglement contour and identify the properties of entanglement spreading. 
We find violation of the linear-light-cone spreading.
(III) We estimate a saturation time of entanglement entropy.
These three observations give reliable evidences of fast scrambler for the present MoCs.
Section V is devoted to conclusion.


\section{Measurement-only Models}
We consider one-dimensional $L$-qubit system (spin-$1/2$ $L$-site system) under periodic boundary conditions, and
study Clifford circuits on that system, where all physical operators and stabilizers are in Pauli group \cite{Gottesman1997,Nielsen_Chuang}. In this work, we ignore the sign and imaginary factors of stabilizers, which give no effects to the following observables we focus on. 
We apply a sequence of projective measurements, understanding that the dynamical application of the measurement corresponds to time-evolution of the circuit system. 
States of the system are described not by many-body wave functions, but by a set of stabilizers, which belong to Paul group and applicable to both pure and mixed states. 
In this work, we focus on pure state dynamics. 
Then, the projective measurement acts to a target state and changes the stabilizer group corresponding to that state 
by replacing anti-commutative stabilizers with the measurement operator \cite{Gottesman1997,Aaronson2004}, inducing nontrivial quantum dynamics.

\begin{figure}[t]
\begin{center} 
\vspace{0.5cm}
\includegraphics[width=8.8cm]{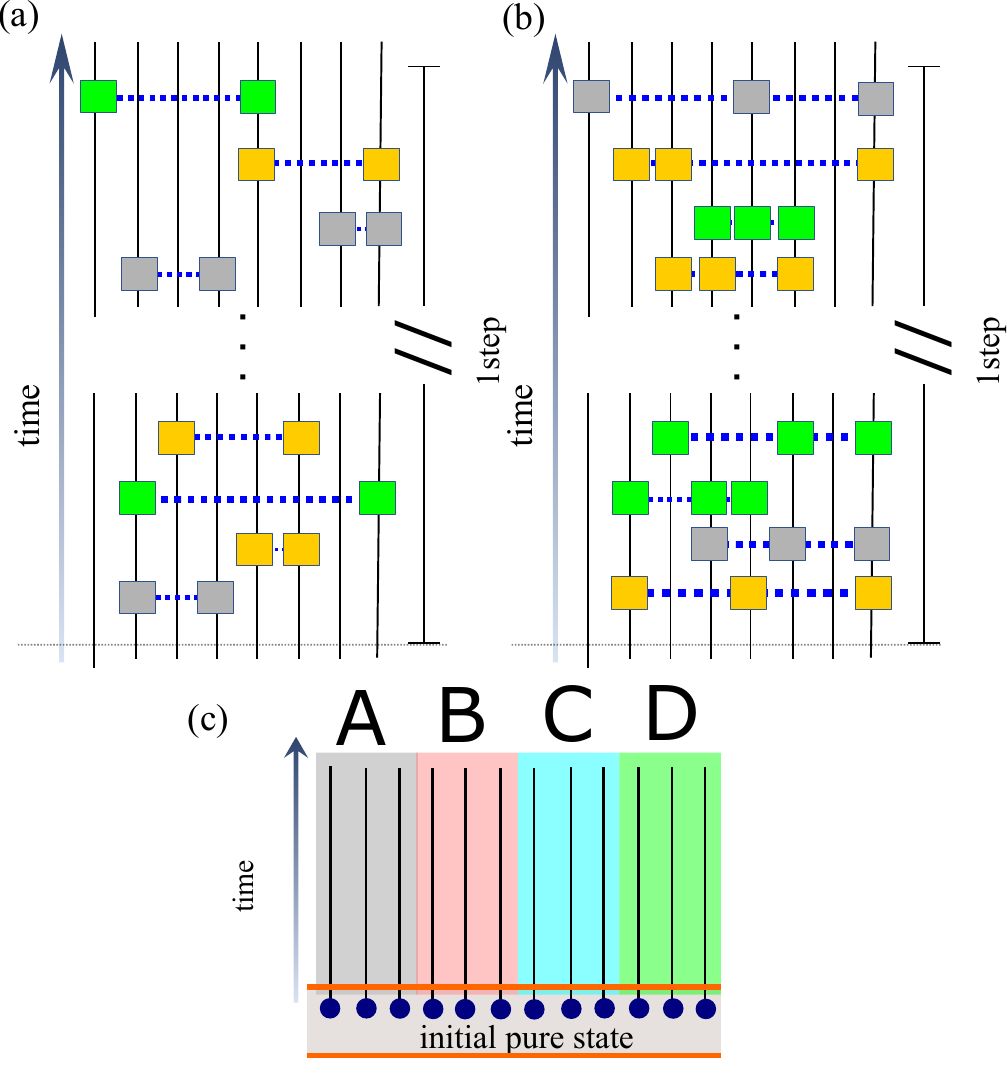}  
\end{center} 
\caption{Schematic image of circuit setting for LR2BM[(a)] and LR3BM[(b)]. 
Unit of time (1-time step) includes $L$-projective measurement blocks. Green, grey, and yellow boxes represent measurement operators Pauli operators, $X$ $Y$,$Z$, respectively.
(c) Schematic image of circuit system and four subsystems ($A,B,C,D$) on one-dimensional qubits for calculation of physical observables. Each subsystem includes $L/4$-qubits. 
In particular, for the TMI we focus on correlation $A$, $B$, and $C$. 
For the negativity and the entanglement entropy, we focus on $A$ and $B$. 
There is no overlap site between these subsystems. Periodic boundary conditions are considered}
\label{Fig1}
\end{figure}

In this work, we consider two distinct types of MoCs, where the circuit is based on one-dimensional $L$-qubits. The spatial sites are labeled by $j=0,1, \cdots, L-1$ for both circuits. 
The first model is a long-range two-body measurement-only model (LR2BM), in which
measurement operators are given by
\begin{eqnarray}
{\hat M}(1,j,r)&\equiv&X_{j}X_{j+r},\\
{\hat M}(2,j,r)&\equiv&Y_{j}Y_{j+r},\\
{\hat M}(3,j,r)&\equiv&Z_jZ_{j+r},
\end{eqnarray}
where $j=0,1,\cdots, L-1$. 
Their projection measurement operators are $P_{\pm}(\sigma,j,r)=\frac{1\pm {\hat M}(\sigma,j,r)}{2}$
with outcome $\pm 1$ and $\sigma=1,2,3$.
We determine the distance $r$ for each measurement with a probability distribution $p(r)\propto r^{\gamma}$ with $\sum_{r}p(r)=1$, 
where $1\leq r \leq L/2$ and $\gamma$ is a (non-positive) controlled parameter in this work. The schematic image of the circuit is shown in Fig.~\ref{Fig1}(a).
Type of measurement operator and site $j$ of each projective measurement are chosen randomly with equal probability. 
Especially, the $r=1$ case is closely related with the projective transverse field Ising model, 
the entanglement phase transition of which was investigated in Ref.~\cite{Lang2020}. 
We shall evaluate saturation values of physical quantities for long-time evolution with the total number of steps $t=20$ in the circuit, 
where the unit of time includes $L$-random projective measurements. 
Most of measurement operators do not commute with each other, and thus, sequential projective measurements induce nontrivial dynamics, 
which can be studied by the classical simulation algorithm \cite{Gottesman1997,Aaronson2004,Nielsen_Chuang}.

The second circuit is a long-range three-body measurement model (LR3BM).
The measurement operators of the model are given by
\begin{eqnarray}
{\hat M}(1,j,r)&\equiv&X_{j}X_{k}X_{j+r},\\
{\hat M}(2,j,r)&\equiv&Y_{j}Y_{k}Y_{j+r},\\
{\hat M}(3,j,r)&\equiv&Z_jZ_{k}Z_{j+r},
\end{eqnarray}
where $j=0,1,\cdots, L-1$, $j+1\leq k \leq j+r-1$. 
Similarly to the case of the LR2BM, we determine the probability of distance-$r$ measurement with the distribution $p(r)\propto r^{\gamma}$ with $\sum_{r}p(r)=1$, 
and measurement type and site $j$ are chosen randomly with equal probability for each measurement.
The intermediate site $k$ also is chosen randomly with equal probability in $j+1\leq k \leq j+r-1$. 
The schematic image of the circuit is shown in Fig.~\ref{Fig1}(b).
For the case $r=2$, the three-body measurement model is similar to the setup in the MoC proposed in Ref.~\cite{Ippoliti2021}. 
In the circuit, some of measurement operators are strongly anti-commutative. 
As in Ref. \cite{Ippoliti2021}, notion of frustration graph is efficient to judge the emergence of volume-law phase. 
As the model has sufficiently high frustration, it is expected to exhibit a volume-law phase,
although how the long-range properties influence the circuit dynamics is a non-trivial problem.  
Therefore in this work, we shall study the effects of the long-rang properties of the measurements in the present circuit. 
We focus on the non-positive range of $\gamma$.


\section{Numerical analysis}
Throughout this work, we employ an efficient stabilizer formalism to simulate the LR2BM and LR3BM,
which plays an important role to study generic Clifford circuits \cite{Gottesman1997,Aaronson2004,Nielsen_Chuang}.
We study the systems with size up to $L=1024$ and focus on a pure state update, 
that is, the initial state is set to the pure product state stabilized by $X_j=+1$ for all $j$. 
In the numerical simulation, we ignore the sign factor of the outcome of the measurement and also the sign and imaginary factor of stabilizers in the update of the stabilizer algorithm. 
This simplicity gives no substantial influence on the calculation of physical quantities of our interest such as entanglement entropy.

\subsection{Target physical quantities}
We first study global structure of the model in terms of steady state in both the LR2BM and LR3BM. 
As varying the parameter $\gamma$ controlling long-range properties, we observe how entanglement of the steady states changes for both the MoCs.

To elucidate properties of the steady states emerging in the circuits, we calculate three physical quantities that are efficiently calculable on Clifford circuits: 
(a) entanglement entropy (b) tripartite mutual information (TMI) and (c) negativity. 

We explain the above three quantities. 
First, we introduce the entanglement entropy for a subsystem $X$ denoted by $S_X$.
In the MoC (by the stabilizer formalism), the entanglement entropy is related to the number of linearly-independent stabilizers in a target subsystem $X$ \cite{Fattal2004,Nahum2017}. 
It is given by $S_X=g_X-L_X$,
where $L_X$ is system size of the subsystem $X$.
$g_X$ is given by $\mathrm{rank}|M_X|$, where 
the matrix $M_X$ is obtained by stacking binary-represented vectors of $L$ stabilizers, which are spatially truncated within $X$ subsystem. The technical aspects are briefly explained in Appendix A.
In the practical numerics, $\mathrm{rank}|M_X|$ is calculated by Gaussian-elimination procedure under binary form \cite{Nielsen_Chuang}.  

Second, let us introduce the TMI. 
We divide the system into four contiguous subsystems denoted by $\{A,B,C,D\}$ as shown in Fig.~\ref{Fig1}(c). 
By using the entanglement entropy, we define the mutual information between $X$ and $Y$ subsystems 
(where $X, Y$ are some elements of the set of the subsystems $\{A,B,C,D\}$, 
and $X\neq Y$); 
$$
I(X:Y)=S_X+S_Y-S_{XY}.
$$
This quantity quantifies the correlation between the subsystems $X$ and $Y$. 
Further, from the mutual information, the TMI for the subsystems $A$, $B$ and $C$ is given by 
$$
I_3(A:B:C)=I(A:B)+I(A:C)-I(A:BC).
$$
The TMI extracts the correlation among the subsystems A, B, and C. 
That is, the TMI can capture the mutual correlation and scrambling (spread of quantum information) emerging by quantum dynamics of 
the system and thus has been broadly employed \cite{Zabalo2020,Sharma2022,Kuriyattil2023}. 
There, the strength of the correlation is given by the magnitude of negativity of the TMI \cite{Hosur2016}. 
In detail, $I_3(A:B:C)$ captures property of information scrambling more precisely than the mutual information 
since the mutual information quantifies correlation between two subsystem, but the TMI quantifies correlation of quantum information (spread of information) among three subsystems. 
If information spreads over the three regimes $A$, $B$, $C$, then the mutual information $I(A;BC)>I(A;B)+I(A;C)$, 
then $I_3 < 0$ and if information gets localized (especially, information in $A$ does not spread across the three regime), 
then $I(A:B)+I(A:C)=I(A:BC)$ is satisfied, then $I_3=0$. 
We shall employ a rescaled TMI $\tilde{I}_3\equiv I_3/{\ln 2}$ in this work.

Note that the TMI is an efficient physical quantity to study the criticality if circuit model under study exhibits a phase transition. 
Since the TMI has minimal finite-size drifts, its calculation predicts the location of the critical point precisely 
with minimal scaling assumptions on small system sizes~\cite{Zabalo2020}.

\begin{figure}[t]
\begin{center} 
\vspace{0.5cm}
\includegraphics[width=8.5cm]{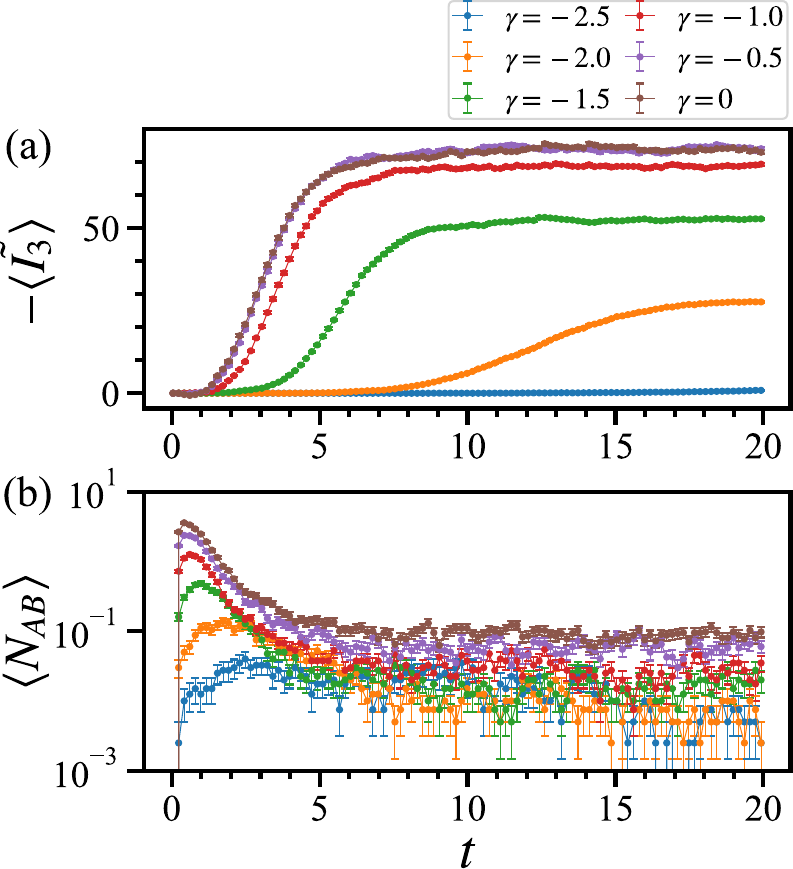}  
\end{center} 
\caption{Late-time dynamics for TMI $\tilde{I}_3$[(a)] and negativity $N_{AB}$[(b)] for LR2BM with system size $L=256$.
For each data, we averaged over $\mathcal{O}(10^2)$ samples.}
\label{Fig2}
\end{figure}

As the third quantity, negativity is considered. 
It can be efficiently calculated in stabilizer formalism, as proposed in Refs. \cite{Shi2020,Sang2021}. 
As in the calculation of the TMI, we consider four contiguous subsystems, $\{A,B,C,D\}$. 
Then, we define the negativity as 
\begin{eqnarray}
N_{AB}=\log_2 ||\rho^{\Gamma_B}_{AB}||,
\end{eqnarray}
where $\rho_{AB}$ is a reduced density matrix for $A\cup B$ subsystem and $\Gamma_B$ denotes the partial transpose of the $B$ subsystem. 
This quantity extracts quantum entanglement  \cite{Groisman2005,MacCormack2021}, 
and intuitively, its value gives the total number of Bell state spanned between $A$ and $B$ subsystems \cite{Sang2021}. The practical way of calculation is mentioned in Appendix A.
The practical calculation shows that the values of $N_{AB}$ are actually small \cite{Shi2020,Sang2021,Sharma2022}, 
but they give useful information complementary to other physical quantities. 

\begin{figure}[t]
\begin{center} 
\vspace{0.5cm}
\includegraphics[width=8.5cm]{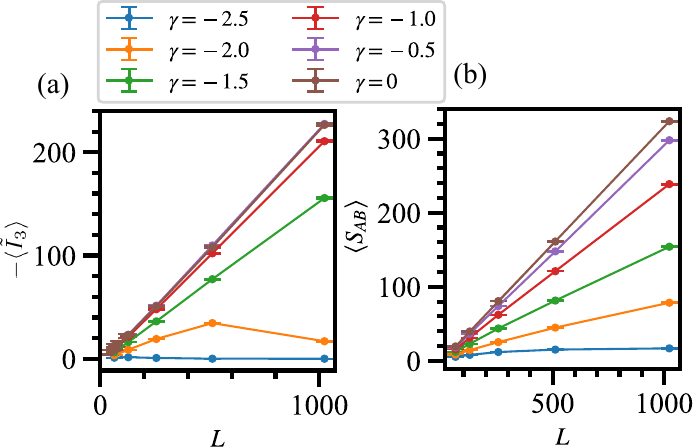}  
\end{center} 
\caption{System-size dependence of TMI [(a)] and half-cut entanglement entropy[(b)] for LR2BM with various values of $\gamma$. 
We averaged over $\mathcal{O}(10^2)-\mathcal{O}(10^3)$ samples for each system size $L$.}
\label{Fig3}
\end{figure}

\subsection{Steady state as varying long-range parameter in LR2BM}

We investigate dynamical and steady state properties of the LR2BM.
We first observe the averaged dynamical behavior observed by the TMI and negativity.
The results for various values of $\gamma$ are shown in Figs.~\ref{Fig2}(a) and ~\ref{Fig2}(b). 
The time evolution of the TMI, $\tilde{I}_3$, exhibits interesting behavior. 
For small negative $\gamma$ (the long-range property is enhanced), the TMI takes large negative values 
while for large negative $\gamma$ (the long-range property is hindered), its magnitude is very small, almost zero for $\gamma=-2.5$. 
This indicates that the long-range property of the measurement, controlled by $\gamma$, gives an essential effect on the dynamics, 
especially on the steady states. 
It leads to the conclusion that strongly-entangled steady states emerge solely by two-body measurements with sufficient long-range properties. 
Late time dynamics of the negativity also supports this physical picture as shown in Fig.~\ref{Fig2}(b). 
That is, the saturation values of the negativity increase as enhancing the long-range measurement by decreasing $|\gamma|$, and 
quantum entanglement gets stronger as approaching the all-to-all case, while for short-range measurement, such correlation is suppressed.
We further note that in relaxation process in early-time periods the negativity exhibits a sudden increase, 
implying that the quantum correlation is generated there.

Next, we observe system-size dependence of the saturation values. 
The saturation values are taken at $t=20$ where the system sufficiently reaches a steady state. 
The calculations of the TMI and half-cut entanglement entropy $S_{AB}$ are shown in Fig.~\ref{Fig3}. 
We find that both data exhibit volume-law scaling such as $S_{AB}\propto L$ and $-\tilde{I}_3\propto L$ for small negative $\gamma$ and all-to-all case,
while for large negative $\gamma$ a kind of subvolume-law scaling appears. 
This implies existence of an entanglement phase transition induced by varying production probability of the long-range measurement. 
We also observe the system-size dependence of the saturation value of the negativity, the results of which are shown in Appendix B. 
The long-range properties of measurement significantly influence on entanglement of steady states.

\subsection{Steady state as varying long-range parameter in LR3BM}
\begin{figure}[t]
\begin{center} 
\vspace{0.5cm}
\includegraphics[width=8.5cm]{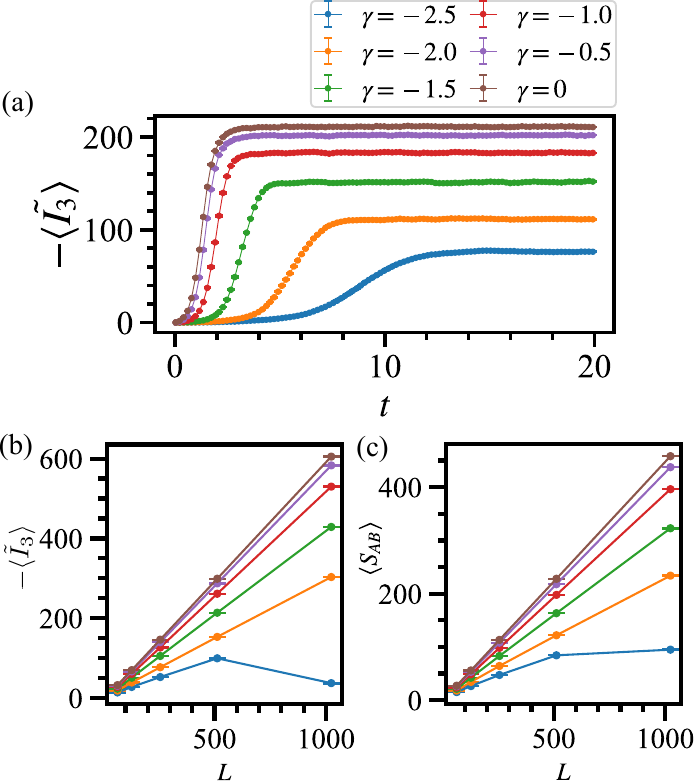}  
\end{center} 
\caption{Obtained results for LR3BM.
(a) Late-time dynamics for TMI $\tilde{I}_3$.
For each data, we averaged over $\mathcal{O}(10^3)$ samples.
(b) and (c): System size dependence of TMI and half-cut entanglement entropy.
For the data (b) and (c), we averaged over $\mathcal{O}(10^2)-\mathcal{O}(10^3)$ samples for each system size $L$.}
\label{Fig4}
\end{figure}

Similar calculation to the case of the LR2BM is carried out for the LR3BM to investigate the dynamical behavior of the TMI.
The $\gamma$-dependent results of the TMI are shown in Fig.~\ref{Fig4}(a). 
For small negative $\gamma$ (the long-range property is enhanced), the TMI has large negative values. 
Even for large negative $\gamma$'s (the long-range property is suppressed), it remains negative. 
This indicates that the long-range property of the measurement generates strong correlations to the dynamics, especially on steady states. 
Contrary to the LR2BM case, even in short-range regime, the correlation remains, implying a volume-law phase survives in the short-range regime. 
This is consistent with the observation in Ref. \cite{Ippoliti2021}. 

As for the negativity, its values under the time evolution keep zero, indicating that the three-body measurement cannot create Bell-pair-like entanglement between $A$ and $B$ subsystems.    

We next observe system-size dependence of saturation values of the TMI and half-cut entanglement entropy $S_{AB}$. 
The saturation value is taken at $t=10$, at which the states reach a steady state. 
The results are shown in Figs.~\ref{Fig4}(b) and (c).
We find the volume-law scaling $S_{AB}\propto L$ and $-\tilde{I}_3\propto L$ for almost the entire range of $\gamma$ up to the all-to-all case. 
For large negative $\gamma$ (short-range measurement), the volume law is weak, and we cannot judge whether it is subvolume law or volume law. We discuss this point later on.

\subsection{Phase transition in LR2BM}
\begin{figure}[t]
\begin{center} 
\vspace{0.5cm}
\includegraphics[width=8.8cm]{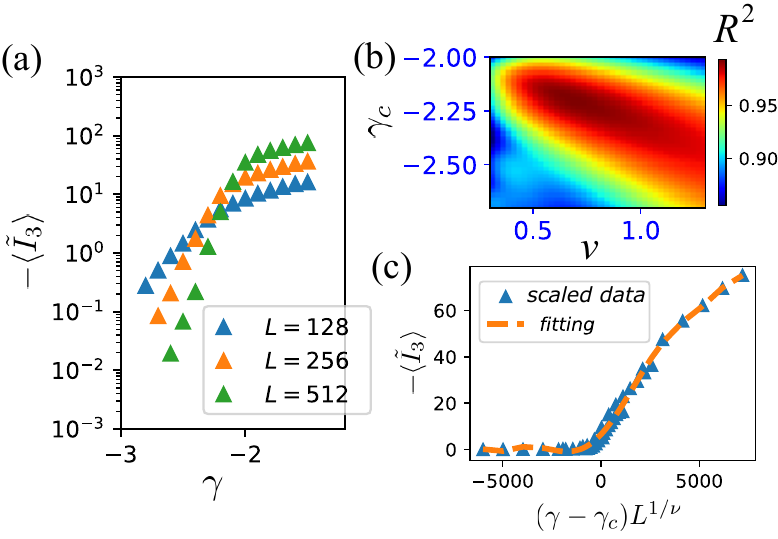}  
\end{center} 
\caption{(a) Different system size data of TMI for LR2BM. (b) 
Heatmap of $R^2$-value. At its maximum, $\gamma_c=-2.21(0)$, $\nu=0.67(6)$. (c) Scaling function for $\gamma_c=-2.21(0)$, $\nu=0.67(6)$. 
We averaged over $10^3, 600, 400$ samples for $L=128$, $256$, $512$, respectively.}
\label{Fig5}
\end{figure}

In the previous subsection, we observed that the long-range property of the projective measurement induces the volume-low phase 
in the LR2BM for small negative $\gamma$, whereas for large negative $\gamma$ (short-range), the subvolume law emerges instead. 
Hence, we expect that there is a phase transition between the above two regimes.
To verify this expectation, we plot the TMI, $\tilde{I}_3$, for different system size in Fig.~\ref{Fig5}(a). 
Here, we find clear data crossing among the different system sizes, indicating the existence of the phase transition.
That is, the enhancement of long-range measurement causes a phase transition.  

To examine the nature of the phase transition, we further carry out a finite size scaling (FSS) analysis. 
We locate the phase transition point of $\gamma$, denoted by $\gamma_c$, and estimate a critical exponent,
by using the obtained (averaged) values of $-\langle \tilde{I}_{3}\rangle$. 
Here, we employ the following scaling ansatz \cite{Ippoliti2021,Kuriyattil2023},
\begin{eqnarray}
-\langle \tilde{I}_{3}\rangle (\gamma, L)=\Psi((\gamma-\gamma_c)L^{1/\nu}),
\end{eqnarray}
where $\Psi$ is a scaling function and $\nu$ is a critical exponent. 
Figures \ref{Fig5}(a), \ref{Fig5}(b) and \ref{Fig5}(c) display the results of the FSS. 
All data of different system sizes obviously collapse into a single curve indicating a genuine phase transition.
By using an optimization procedure, we determine the values of $\gamma_c$ and $\nu$. 
From the maximum $R^2$-value, we obtain $\gamma_c=-2.21(0)$ and $\nu=0.67(6)$, with the maximal $R^2=0.993(0)$. 

The above FSS analysis indicates that the long-range properties of the LR2BM induce the entanglement phase transition. 
To corroborate this conclusion, we also study the behavior of the saturation values of the negativity. 
The result is shown in Appendix B, which supports the existence of the phase transition.

Here, a remark is in order. 
As we showed, the LR2BM does not exhibit volume-law entanglement in the vicinity of the nearest-neighbor limit.
This is consistent with the observation using frustration graph in Ref.~\cite{Ippoliti2021}.
However, the present work shows that the long-range LR2BM exhibits volume-law entanglement phase.
This result, therefore, implies that the argument of the frustration graph should be used with caution for circuits including long-range measurements.

In the LR3BM for the whole range of $\gamma$, steady states exhibit the volume-law entanglement. 
For the behavior of the TMI, the crossing among different system size data does not occur. 
That is, even for the limit $\gamma\to -\infty$, the LR3BM stays in the volume-law phase.
This result is consistent with the observation in the previous study \cite{Ippoliti2021}. 
This numerical details are given in Appendix C.

\begin{figure}[t]
\begin{center} 
\vspace{0.5cm}
\includegraphics[width=7.5cm]{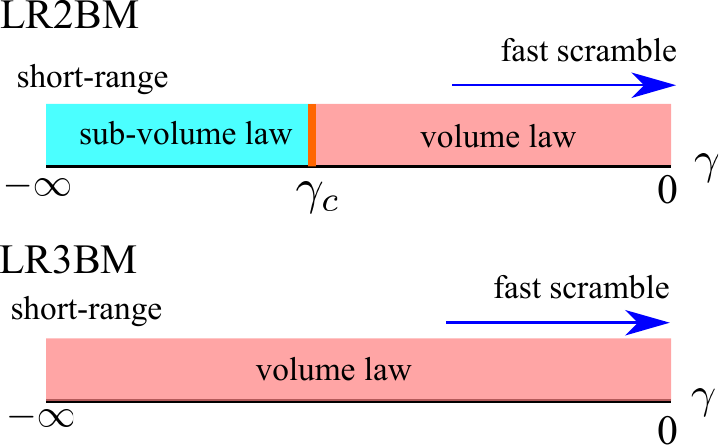}  
\end{center} 
\caption{Steady state entanglement phase diagrams for LR2BM and LR3BM. The phase of LR2BM is clearly separated into volume and subvolume law phases.
For the neighboring three-body cluster limit in the lower phase diagram for LR3BM, 
the volume law phase remains, consistent with the prediction of the previous study \cite{Ippoliti2021}.}
\label{Fig6}
\end{figure}

We show the summary of the global phase diagram for both the LR2BM and LR3BM in Fig.~\ref{Fig6}. 
In the following, we study properties of entanglement growth in the volume-law phase in detail. 

\begin{figure}[t]
\begin{center} 
\vspace{0.5cm}
\includegraphics[width=8.5cm]{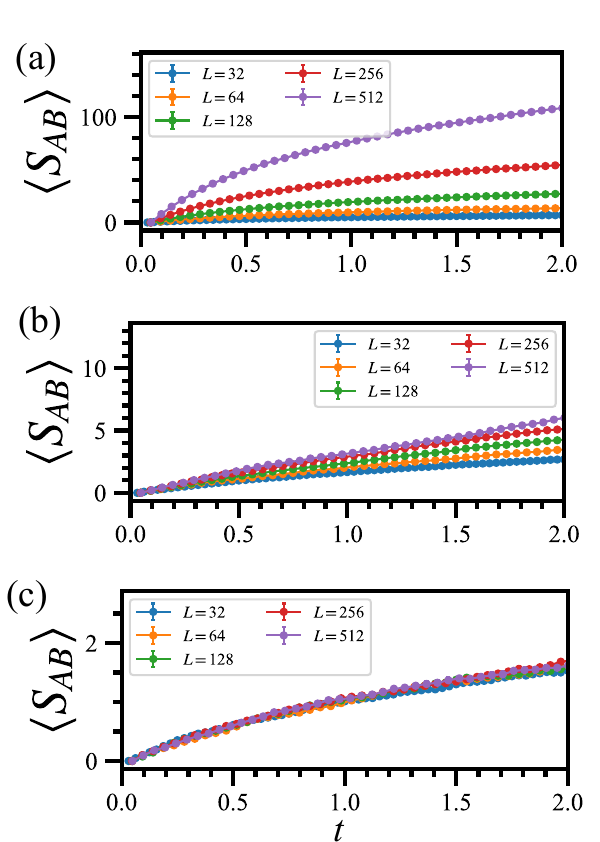}  
\end{center} 
\caption{Intermediate dynamics of entanglement entropy for half subsystem for LR2BM. 
The time evolution is up to $t=2$.
We plot three long range cases, (a) $\gamma=0$, (b) $\gamma=-2$, (c) $\gamma=-3$. The case (a) and (b) are in volume law phase. The other (c) is in subvolume law phase. 
For all data, we averaged over $\mathcal{O}(10^2)-\mathcal{O}(10^3)$ samples for each system size $L$.}
\label{Fig7}
\end{figure}

\begin{figure}[t]
\begin{center} 
\vspace{0.5cm}
\includegraphics[width=8.8cm]{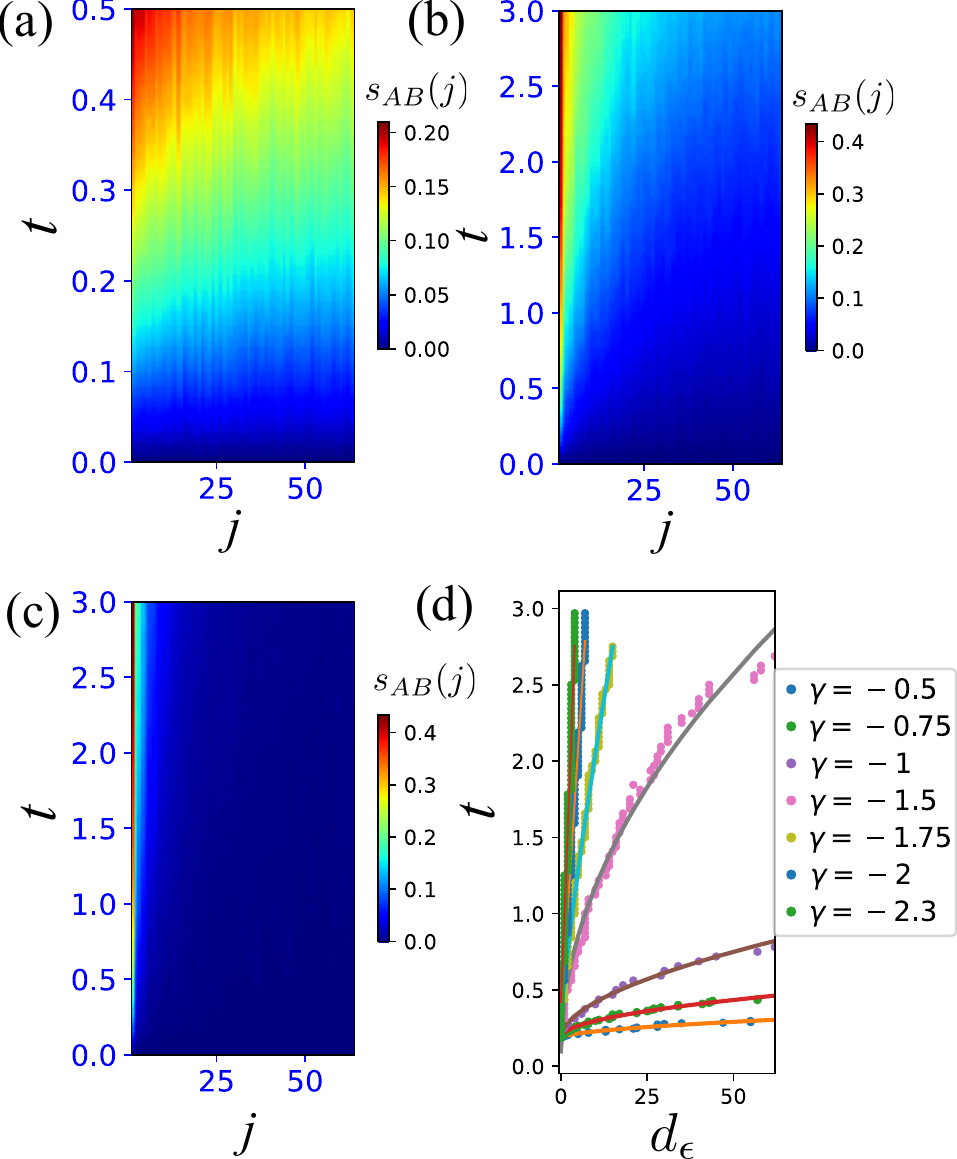}  
\end{center} 
\caption{Entanglement contour $s_{AB}(j)$ with $j=0,\cdots,L_{AB}/2-1$ for $\gamma=-0.5$ [(a)], $\gamma=-1.5$ [(b)] and $\gamma=-2.5$ [(c)]. 
We set $L=256$, $L_{AB}=128$.
All data are obtained by averaging over $\mathcal{O}(10^4)$ samples. 
(d)Wave front $d_{0.1}(t)$ for various $\gamma$ and fitting curve is set $t=a_0 d_{0.1}^{a_1}(t)+a_2$, where $a_\ell$'s ($\ell=0,1,2$) are fitting coefficients.}
\label{Fig8}
\end{figure}
\begin{figure}[t]
\begin{center} 
\vspace{0.5cm}
\includegraphics[width=8.5cm]{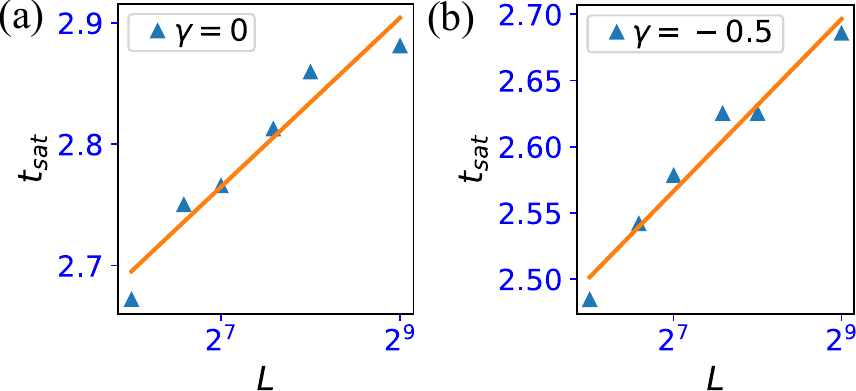}  
\end{center} 
\caption{Saturation time $t_{\rm sat}$ vs system size $L$ for half-cut entanglement entropy for LR2BM:
(a) $\gamma=0$ data with $\alpha=0.8$, (b) $\gamma=-0.5$ with $\alpha=0.78$. 
For all data, we averaged over $\mathcal{O}(10^2)-\mathcal{O}(10^3)$ samples for each system size $L$.}
\label{Fig9}
\end{figure}

\section{Detailed analysis of entanglement growth}
We observed the volume-law phase for both the LR2BM and LR3BM. 
The LR2BM exhibits a phase transition from subvolume-law to volume-law phase as increasing $\gamma$.
On the other hand, the LR3BM is in the volume-law phase for the whole parameter range of $\gamma$.

Here we investigate whether or not the long-range measurements give significant effects on dynamics in the MoCs. 
In particular for the volume-law entanglement phase, we are interested how scrambling behavior changes on varying the long-range parameter $\gamma$. 
Intuitively, it is expected that the enhancement of the long-range properties of the projective measurement can generate trend of fast scrambling. 
Similar possibility was examined for a spin chain with long-range or all-to-all coupling \cite{Bentsen2020,Belyansky2020,Yao2016,Li2020}.
This issue has been extensively investigated in hybrid random-unitary circuits \cite{Hashizume2022,Hashizume2022_2,Kuriyattil2023},
but study on this issue is still lacking for MoCs. 
Here, we systematically investigate the above problem by observing various physical quantities, which are to be efficiently calculated in the stabilizer formalism. 

In what follows, we numerically calculate the following three observables for both the LR2BM and LR3BM: 
(I) time evolution of half-cut entanglement entropy $S_{AB}$, 
(II) entanglement contour and its wave front and (III) system-size dependence of the saturation time of the half-cut entanglement entropy $S_{AB}$.

In particular, the observable (II) gives an important insight about entanglement growth. 
The entanglement contour proposed in \cite{Chen2014,MacCormack2021,MacCormack2020} provides a clear picture of a spatial profile of entanglement.  
We focus on entanglement entropy of the subsystem $A\cup B$ as shown in Fig.~\ref{Fig1}(c) (half-cut subsystem).
The entanglement contour \cite{Chen2014,MacCormack2021} is defined as
\begin{eqnarray}
s_{AB}(X_j)&=&\frac{1}{2}[S(X_j|X_0\cup X_1\cdots X_{j-1})\nonumber\\
&&+S(X_j|X_{j+1}\cup X_2\cdots X_{L_{AB}-1})],
\label{s_contour}
\end{eqnarray}
and here
\begin{eqnarray}
S(X|Y)&\equiv&S_{X\cup Y}-S_Y,
\end{eqnarray}
where $S_{X\cup Y}(=S_{XY})$ is entanglement entropy of the subsystem $X$ and $Y$ within the subsystem $A\cup B$, $j$ is a site label, 
$j=0,1,\cdots L_{AB}-1$ and $L_{AB}$ is the number of qubits in $A\cup B$ subsystem.
Methods of the numerical calculation of $s_{AB}(X_j)$ are explained in Appendix D.
We set the subsystems $\{X_j\}$ to smallest ones, that is, a single site in $A\cup B$-subsystem, $X_j \to j$. 
Then, $0\leq s_{AB}(j)\leq 1$.

From the results of the entanglement contour, we can define its wave front as
\begin{eqnarray}
d_\epsilon(t)=\max [\{j| j\in A:  s_{AB}(j)>\epsilon\}].
\label{depsilon}
\end{eqnarray}
Time evolution of the wave front gives us the insight into the spatial spreading of entanglement and correlation 
in the system, and clarifies spreading behavior such as linear light cones (Lieb-Robinson bound), etc. 
In a recent study \cite{Sang2022}, similar quantity to $d_\epsilon(t)$ was estimated to extract the behavior of the linear light-cone spreading.
In this work, for the practical calculation of $d_\epsilon(t)$, we set $\epsilon=0.1$ in Eq.~(\ref{depsilon}),
but the qualitative results are the same for other values of $\epsilon$. 

Turning to the observable (III), we estimate the saturation time of half-cut entanglement entropy $S_{AB}$. 
We denote the saturation time by $t_{sat}$. 
As study in Refs.~\cite{Sekino2008,Maldacena2016} shows, when fast growth of entanglement takes place in the system, 
the saturation time, which is defined in terms of out-of-time ordered correlation, exhibits the system-size dependence such as
$t_{sat}\propto \log L$. 
For different definition of saturation time by using quantities such as half-cut entanglement entropy $S_{AB}$, mutual information and TMI,
a similar relationship between the saturation time and system size holds in systems of fast scrambler.
In fact, study using mutual information and entanglement entropy has been reported in Ref.~\cite{Hashizume2022_2}. 
Here, a comment is in order.
Besides the bound of saturation time under fast scrambling such as $t_{sat}\propto \log L$, 
other possibilities for the relationship between saturation time and system size have been suggested in Refs.~\cite{Bentsen2019,Lucas2019}.
In this work, we estimate the saturation time of the half-cut entanglement entropy $S_{AB}$ and examine its system-size dependence. 

\subsection{Results of LR2BM}
Let us show the numerical results of the observation (I), (II) and (III) for the LR2BM. 

\noindent \underline{Observable (I):} 
The intermediate-time dynamics of the half-subsystem entanglement entropy $S_{AB}$ for various values of $\gamma$ 
and system sizes are displayed in Fig.\ref{Fig7}. 
We focus on the intermediate-time regime, before the saturation emerges. 
Here, we observe that in Fig.~\ref{Fig7}(a), the all-to-all limit exhibits large system-size dependence of the entanglement entropy in the intermediate times. 
This seems a signal of the fast scrambling as discussed in Ref.~\cite{Belyansky2020}. 
That is, the increasing rate of $S_{AB}$ gets larger as the system size increases. 
This tendency remains even for finite negative $\gamma$'s. 
As observed in Fig.~\ref{Fig7}(b) for $\gamma=-2$, the increasing rate again gets larger as the system size increases. 
The cases (a) and (b) exhibit qualitatively the same dynamical phenomenon, which is consistent with the observation that both cases are in the volume-law phase. 
Hence, we expect that the volume-law entanglement phase in the LR2BM possess fast scrambling property.
But in the case $\gamma=-3$ shown in Fig.~\ref{Fig7}(c), such a behavior does not appear, that is, no system size dependence. 
We expect that scrambling does not occur in this phase. 

\noindent\underline{Observation (II):} 
We next observe the entanglement contour for half-subsystem denoted by $A\cup B$ ($AB$).
The numerical results for the typical vales of $\gamma$ are shown in Figs.~\ref{Fig8}(a)-(c). 
We see that the spreading image of the case $\gamma=-0.5$ is significantly different from linear light cones as shown in Fig.~\ref{Fig8}(a).
The strong long-range property of 2-body measurements dramatically changes the linear-light-cone propagation of entanglement entropy.
This behavior survives as decreasing $\gamma$ (long-range property) to $\gamma=-1.5$, as seen in Fig.~\ref{Fig8}(b). 
On the other hand for further short-range regime, such a non-linear-light cone behavior disappears as shown in Fig.~\ref{Fig8}(c). 
At least, we expect that for negative small $\gamma$ (long-range regime), fast scrambling phenomenon occurs. 
This fast scrambling phenomenon is expected to be universal in the volume-law phase in the LR2BM.

Furthermore, we extract the wave front from the data of the entanglement contour for various $\gamma$'s. 
The obtained result is shown in Fig.~\ref{Fig8}(d). 
We find that in the case of the volume-law regime, the curve of $d_\epsilon(t)$ is not linear but exhibits power-law behavior, 
while in the vicinity of the phase transition $\gamma=-2.3$, the curve of $d_\epsilon(t)$ is close to linear. 
We examine the power-law behavior of $d_\epsilon(t)$ by using the following fitting curve, 
$
t=a_0 d_{0.1}^{a_1}(t)+a_2.
$
From the data of $d_{0.1}^{a_1}(t)$, we estimate the coefficients $(a_0, a_1,a_2)$.
In particular, our interest is the exponent $a_1$. 
The obtained results are the following: 
For $\gamma=-0.5,-0.75 -1,-1.5,-1.75, -2,-2.3$, $a_{1}=0.56(9),0.50(2),0.56(0),0.51(6),0.82(5),1.01(8),1.05(4)$, respectively. 
From the above data, we expect that $a_1$ is universal in the volume-law regime, roughly $a_1\sim 0.5$, 
that is, the entanglement spreads as $d_{\epsilon}\propto t^{2}$.
On the other hand in the vicinity of the phase transition point and also in subvolume-law regime, $a_1\sim 1$, 
that is, the entanglement spreads as $d_{\epsilon}\propto t$, linear light cones. 
We think that in the strong long-range measurement regime with the volume-law entanglement, the correlation and entanglement spread as $t^2$ 
implying super-ballistic, a signature of fast scrambling in contrast to the linear-light-cone propagation. 

We further comment that for near the all-to-all case of the LR2BM such as $\gamma=-0.5$ and $-0.75$, 
an exponential fitting can be applied as $d_{\epsilon}(t)=b_0\exp[b_1 t]+b_2$, where $(b_0,b_1,b_2)$ are fitting parameters. 
The obtained results are: for $\gamma=-0.5$, $b_1=14.8(0)$ and for $\gamma=-0.75$, $b_1=9.73(2)$. 
(Inversely, implying $t\propto \log d_{\epsilon}$.)
Even though we compare $R^2$-values of both power-law and exponential fittings, we cannot judge which one is better, 
as $R^2>0.95$ for both of them. 
Therefore, at least, we conclude that the linear-light-cone picture is significantly altered by the long-range measurement.

\begin{figure}[t]
\begin{center} 
\vspace{0.5cm}
\includegraphics[width=9cm]{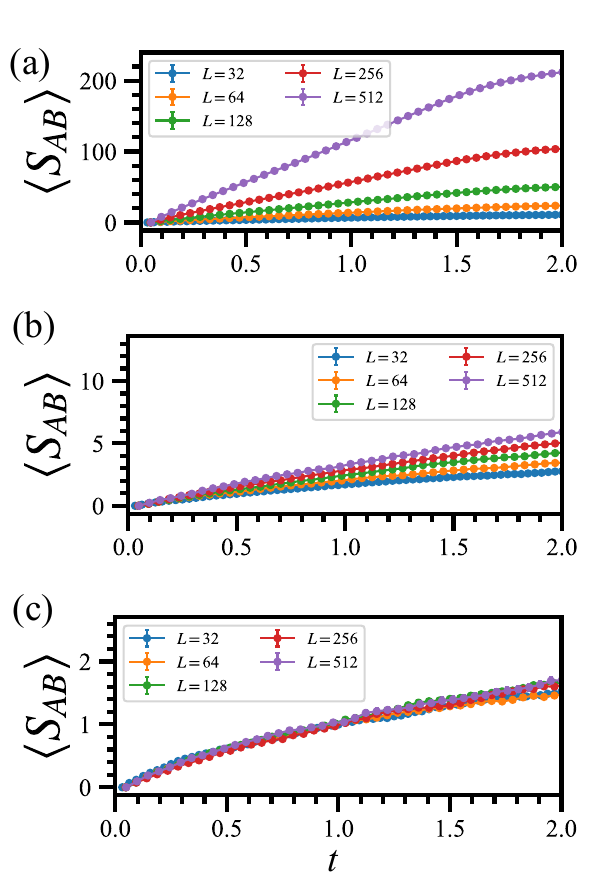}  
\end{center} 
\caption{Intermediate dynamics of entanglement entropy for half subsystem for LR3BM. 
The time evolution is up to $t=2$.
We plot three long range cases, (a) $\gamma=0$, (b) $\gamma=-2$, (c) $\gamma=-3$. 
The all cases are in volume law phase. 
For all data, we averaged over $\mathcal{O}(10^2)-\mathcal{O}(10^3)$ samples for each system size $L$.}
\label{Fig10}
\end{figure}

\noindent\underline{Observation (III):}
Finally in this subsection, we observe the saturation time vs system size $L$ in the half-cut entanglement entropy $S_{AB}$. 
Practically for all cases in the MoC, the half-cut entanglement entropy $S_{AB}$ saturates in the regime $5\le t \le 12$. 
Then, by employing the time-averaged values of the half-cut entanglement entropy $S_{AB}$, we determine the saturation time as
$t_{\rm sat}=\min[\{t |t\ge \alpha \bar{E}\}]$, where $\bar{E}$ is the time-averaged value of the half-cut entanglement entropy $S_{AB}$.
Here, we focus on the time regime $5\le t \le 12$ and $\alpha$ is a tuning parameter. 
Note that in the estimation of $t_{\rm sat}$, 
we do not concern the error of $S_{AB}$ since the standard error in averaging $S_{AB}$ is negligibly small as shown in Fig.~\ref{Fig7}. 

By the practical calculation, we obtain very important results for relation between saturation time $t_{\rm sat}$ and system size $L$ plotted in Fig.~\ref{Fig9}
for $\gamma=0$ and $\gamma=-0.5$. 
We clearly find $t_{\rm sat}\propto \log_2$ L in this MoC. 
This result is similar to that obtained by studying the out-of-time-ordered correlator in Refs. \cite{Sekino2008,Li2020,Belyansky2020},
and indicates the fast scrambling takes place in the long-range enhanced regime of the LR2BM.

\begin{figure}[t]
\begin{center} 
\vspace{0.5cm}
\includegraphics[width=9cm]{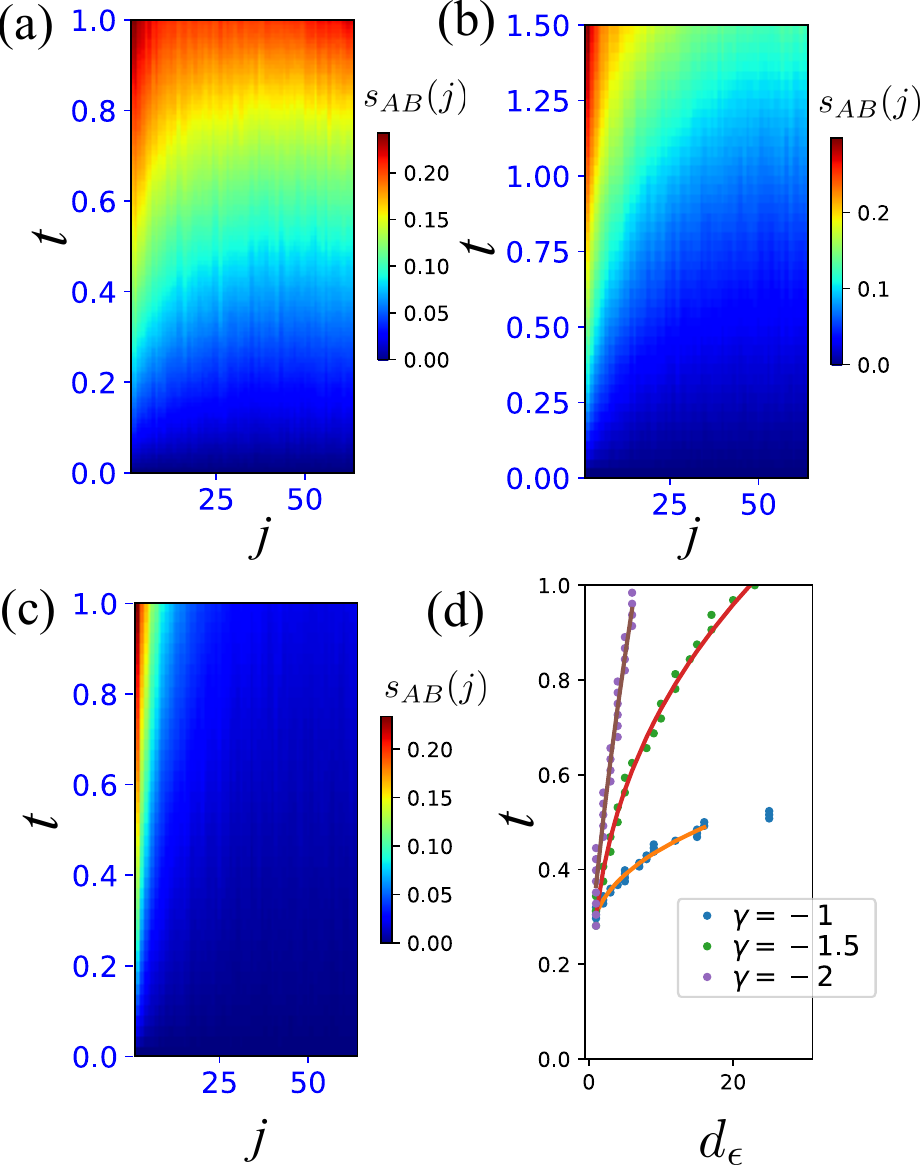}  
\end{center} 
\caption{Entanglement contour $s_{AB}(j)$ with $j=0,\cdots,L_{AB}/2-1$ for $\gamma=-1$ [(a)], $\gamma=-1.5$ [(b)] and $\gamma=-2$ [(c)]. 
We set $L=256$, $L_{AB}=128$.
All data are obtained by averaging over $\mathcal{O}(10^4)$ samples. 
(d) Wave front $d_{0.1}(t)$ for various $\gamma$ and the fitting curve for early time dynamics is set 
$t=a_0 d_{0.1}^{a_1}(t)+a_2$ where $a_\ell$'s ($\ell=0,1,2$) are fitting coefficients.}
\label{Fig11}
\end{figure}

\subsection{Results of LR3BM}
Let us turn to the numerical results of the LR3BM. 
\underline{Observation (I):} 
The result of the half-cut entanglement entropy $S_{AB}$ for various long-range parameter $\gamma$ with different system sizes are displayed in Fig.\ref{Fig10}. 
The same tendency to the case of the LR2BM occurs. 
As shown in Fig.~\ref{Fig10}(a), the all-to-all limit exhibits large system-size dependence of the entanglement entropy in the intermediate-time period. 
Fig.~\ref{Fig10}(b) exhibits similar dynamical behavior, which indicates that both cases are in the same volume-law phase. 
However even in the volume-law phase, the case $\gamma=-3$ shown in Fig.~\ref{Fig10}(c) exhibits no significant system-size dependence. 
We expect that scrambling is weak in the limit of the neighboring three-body cluster measurement.

\noindent\underline{Observation (II):} 
We next display the entanglement contour for half-subsystem $A\cup B$($AB$).
The numerical results for the typical $\gamma$'s are shown in Figs.~\ref{Fig11}(a)-(c). 
The data (a) and (b) show significant nonlinear propagation, while the short-range case (c) exhibits linear light-cone-like propagation.

Furthermore, we observe the behavior of the wave front from the data of the entanglement contour for various $\gamma$'s. 
The obtained results are shown in Fig.~\ref{Fig11}(d).
From the data of $d_{0.1}^{a_1}(t)$, we estimate the coefficients $(a_0,a_1,a_2)$. 
Especially, the data of the wave front of early-time dynamics for $\gamma=-1,-1.5, -2$ give estimated values $a^{1}=0.38(3),0.33(1),0.75(1)$, respectively. 
All value are smaller than $1$, implying the breakdown of linear-light-cone picture. 
Therefore, we expect that as enhancing long-range property of measurement, the correlation propagation obeys $d_{\epsilon}\propto t^{3}$, 
implying super-ballistic.
while for short-range regime, the propagation approaches linear light cones. 

\begin{figure}[t]
\begin{center} 
\vspace{0.5cm}
\includegraphics[width=8.5cm]{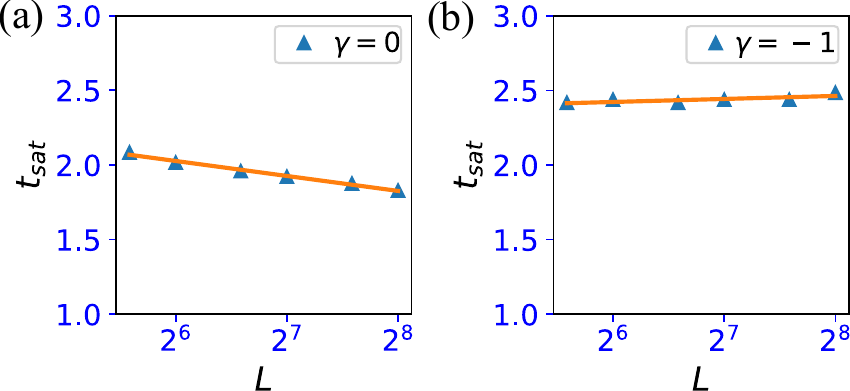}  
\end{center} 
\caption{Saturation time $t_{\rm sat}$ vs system size $L$ for half-cut entanglement entropy $S_{AB}$. 
(a) $\gamma=0$ data with $\alpha=0.9$, (b) $\gamma=-1$ with $\alpha=0.9$. 
For all data, we averaged over $\mathcal{O}(10^2)-\mathcal{O}(10^3)$ samples for each system size $L$.}
\label{Fig12}
\end{figure}

\noindent\underline{Observation (III):}
Finally in this subsection, relation between the saturation time and system size $L$ in the half-cut entanglement entropy $S_{AB}$ 
is observed as in the case of the LR2BM.  
The results are plot in Fig.~\ref{Fig12} for $\alpha=0.9$. 
For the long-range measurement cases, $t_{\rm sat}$ has only small system-size dependence.
In particular for $\gamma=0$ (all-to-all case), a tiny inverse (negative) logarithmic dependence on $L$ exits, whereas for $\gamma=-1$, 
there is almost no system-size dependence. 
We emphasize that this behavior clearly deviates from the linear-light-cone behavior and indicate the fast scrambling in the LR3BM. 
Similar violation of linear light cones has been reported in a random unitary circuit \cite{Bentsen2019}, 
where the saturation time has no system-size dependence, called ultra-fast scrambling.

\section{Conclusion}
In this work, we numerically studied two MoC models called LR2BM and LR3BM. 
We found both the models exhibit volume-law phase as steady state, the presence of which is preserved by the long-range multi-body projective measurements. 
In particular, we found a long-range measurement induces phase transition without any unitary gates in the LR2BM, 
while the LR3BM possesses volume-law phase in the entire parameter range.
We further studied dynamics, especially, scrambling properties for both the LR2BM and LR3BM in detail. 
We numerically extracted evidences of fast scrambler from behaviors of various physical quantities. 
Indeed, MoC model has potential to exhibit fast scrambling. 
In the volume-law phase, the long-range measurement gives strong influence on entanglement growth. 
From the observation of the entanglement contour, a deformed wave is found, which is clearly different from a typical entanglement spreading, linear light cones 
(indicating super-ballistic spreading). 
We also identified that the saturation time of entanglement entropy is indeed faster than that of conventional light cones. 
These intensive numerical studies indicate that long-range multi-body measurements generate fast scrambler-like behaviors.

The present work is the first step toward discovery of various systems exhibiting fast scrambling in quantum circuit models with strong measurements. 
The study of scrambling property for various quantum circuit geometry such as star graph and tree-like one \cite{Bentsen2019,Bentsen2020} 
and strength of measurement \cite{Sang2022} will be a promised future direction of study.

\section*{Acknowledgements}
This work is supported by JSPS KAKENHI Grant Number JP23K13026 (Y.K.) and JP23KJ0360(T.O.). 

\appendix


\section*{Appendix A: Practical calculation of entanglement entropy and negativity}
\noindent \underline {Entanglement entropy}: 
In $L$-site pure and unique system, the system is described by $L$-stabilizers. 
Each stabilizers $g_{\ell}$ without sign and imaginary factor is represented by a $2L$ binary vector ($\ell=0,1,\cdots,L-1$):
\begin{eqnarray}
&&g_{\ell}=\prod^{L-1}_{j=0}(X_{j})^{w^x_j}\prod^{L-1}_{j=0}(Z_{j})^{w^z_j},\\
&&\longrightarrow w_{\ell}=(w^x_0,\cdots, w^x_{L-1} | w^z_0,\cdots, w^z_{L-1}),
\end{eqnarray}
where $w^{\beta}_{j}=0$ or $1$ and we call $w_{\ell}$ stabilizer binary vector.
Here, we consider a subsystem $X$, which includes the site $j=0,\cdots, k$ ($< L-1$), and calculate the entanglement entropy for the subsystem $X$. 
To this end, we first truncate each stabilizer binary vector as 
\begin{eqnarray}
w_{\ell} \longrightarrow  w^X_{\ell}=(w^x_0,\cdots, w^x_{k} | w^z_0,\cdots, w^z_{k}),
\end{eqnarray}
then we create a matrix $M_X$ by stacking the $L$-truncated binary vectors of $L$ stabilizers,
$$
M_X=(w^X_{0},w^X_{1},\cdots,w^X_{L-1})^t.
$$
For this matrix $M_X$, we calculate
$\mathrm{rank}|M_X|$.
The value $L-\mathrm{rank}|M_X|$ is equal to the number of linearly-independent stabilizers within the subsystem $X$ \cite{Fattal2004}. 
As shown in Refs.~\cite{Nahum2017,Lavasani2021}, the entanglement entropy $S_X$ is related to 
$\mathrm{rank}|M_X|$ as 
$S_{X}=\mathrm{rank}|M_X|-L_X$, where $L_X$ is the number of total site of the subsystem $X$.\\

\noindent\underline {Negativity}: The original methods calculating the negativity and its details can be seen in Refs.~\cite{Sang2021_v2,Shi2020}. 
In this Appendix, we explain briefly the methods that we employed in this work.
As explained in Refs.~\cite{Sang2021_v2,Shi2020,Sharma2022}, the negativity is defined by 
\begin{eqnarray}
N_{AB}=\frac{1}{2}\mathrm{rank}J,
\label{J_rep}
\end{eqnarray}
where $J$ is a $m_{AB}\times m_{AB}$ matrix, and $m_{AB}$ is defined as we explain in the following. 
Here, we focus on how to construct the matrix $J$.
First, we consider the partition such as in Fig.~\ref{Fig1}(c). 
Then from the $L$-stabilizers $w_{\ell}$ we create a stacking matrix $M$
$$
M=(w_{0},w_{1},\cdots,w_{L-1})^t.
$$
For this matrix $M$, we transform the matrix $M$ into a standard form \cite{Nielsen_Chuang} denoted by $M^s$, where the spatial index are not transformed, 
that is, the columns are fixed. 
From $M^s$, we extract $m_{AB}$ specific independent generators of stabilizers, which have spatial support within $AB$-subsystem. 
We denote them by $g^{AB}_{\ell}$ with $\ell=0,1,\cdots, m_{AB}-1$. 
Next, we manipulate $g^{AB}_{\ell}$'s by truncating each stabilizer binary vector as 
\begin{eqnarray}
g^{AB}_{\ell} \longrightarrow  g^A_{\ell}=(g^x_0,\cdots, g^x_{k} | g^z_0,\cdots, g^z_{k}),
\end{eqnarray} 
where only binary components within the subsystem $A$ (here labeled by $0,\cdots, k$) remains. 
Finally, by using the $m_{AB}$-binary stabilizers $g^A_{\ell}$, we construct the commutator matrix:
\begin{eqnarray}
(J)_{ij}=
\begin{cases}
1 & \mbox{if}\:\: \{g^{A}_{i},g^{A}_j\}=0\\
0 & \mbox{if}\:\: [ g^{A}_{i},g^{A}_j]=0
\end{cases}
,
\end{eqnarray}
where $\{g^{A}_{i},g^{A}_j\}=0$ means that the truncated stabilizer generators $g^{A}_{i}$ and $g^{A}_j$ are anti-commuting,
and similarly $[g^{A}_{i},g^{A}_j]=0$ means that the truncated stabilizer generators $g^{A}_{i}$ and $g^{A}_j$ are commuting.
By this manipulation, we obtain the binary $m_{AB}\times m_{AB}$ matrix $J$. By using Eq.~(\ref{J_rep}), we obtain the negativity $N_{AB}$.

\section*{Appendix B: $\gamma$-dependence of saturation value of negativity}
\begin{figure}[t]
\begin{center} 
\vspace{0.5cm}
\includegraphics[width=7.5cm]{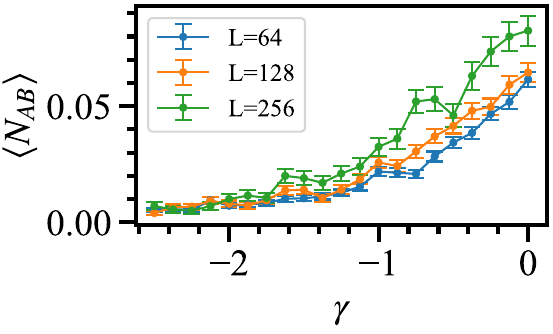}  
\end{center} 
\caption{Saturation value of $\langle N_{AB}\rangle$ as a function of $\gamma$ 
for $L=64$, $128$, and $256$ with $6000$, $4000$, and $2000$ samples, respectively.
}
\label{Fig13}
\end{figure}

We show the $\gamma$-dependence of saturation value of negativity in Fig.~\ref{Fig13}. 
The values of $N_{AB}$ are entirely small, at most $\mathcal{O}(10^{-1})$. 
$N_{AB}$ increases for negative small $\gamma$ (long-range measurement regime). 
This indicates that for negative small $\gamma$, quantum correlation is enhanced and the result supports the existence of volume-law phase in the LR2BM. 
Due to the small values of $N_{AB}$, it is difficult to identify any single crossing point among different system size data. 
Thus, phase transition behavior cannot be extracted. 
More detailed study is a future problem.

\section*{Appendix C: No phase transition behavior of TMI in LR3BM}
\begin{figure}[t]
\begin{center} 
\vspace{0.5cm}
\includegraphics[width=7.5cm]{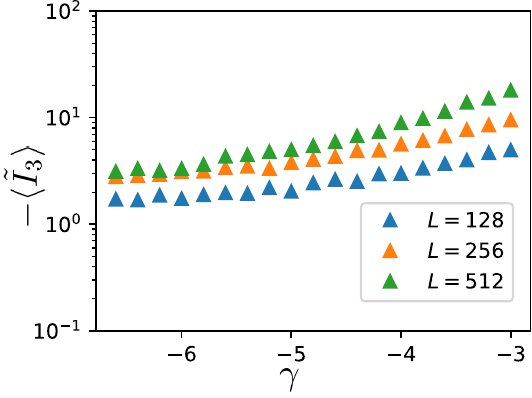}  
\end{center} 
\caption{Behavior of TMI $I_3$ for negative large $\gamma$ (strong short-range cases).
We averaged over $1000, 600, 400$ samples for $L=128$, $256$, $512$, respectively.
.}
\label{Fig14}
\end{figure}

We show the behavior of the TMI as varying $\gamma$ in the LR3BM. 
The numerical result is plotted in Fig.~\ref{Fig14}. 
There is no crossing among the lines with different system sizes. 
We expect that even under the limit $\gamma\to -\infty$, the crossing does not occur and volume-law phase remains up to 
the neighboring three-body measurement case, which has been studied in \cite{Ippoliti2021}.  

\section*{Appendix D: Efficient method for calculating entanglement contour}
In this Appendix, we comment on a more efficient method for calculating the entanglement contour $s_{AB}$ compared with the direct calculation.
The first term of the entanglement contour $\frac{1}{2}S(X_j|X_0\cup X_1\cdots X_{j-1})$ (see Eq.~(\ref{s_contour})) 
consists of $S_{X_0\cup X_1\cdots X_{j}}$ and $S_{X_0\cup X_1\cdots X_{j-1}}$.
Thus, calculating $\frac{1}{2}S(X_j|X_0\cup X_1\cdots X_{j-1})$ for all $j$ requires calculating the $S_{X_0\cup X_1\cdots X_{j-1}}$ for all $j$.
This calculation mostly consists of evaluating the rank of a binary representation of the stabilizer set (cf.~Ref.~\cite{Sang2021_v2,Shi2020}).
We now describe the method that reduces the calculation of $S_{X_0\cup X_1\cdots X_{j}}$ for all $j$ to a single calculation of $S_{X_0\cup X_1\cdots X_{L_{AB-1}}}$.
To explain this efficient method, we introduce a matrix $M_j$ and the operation of matrix horizontal concatenation.
The matrix $M_j$ corresponds to the binary representation of the stabilizer set for site $j$ and is an $L\times 2$ matrix over $F_2$.
Matrix horizontal concatenation is an operation to create a new $N\times (M_A+M_B)$ matrix $(AB)$ from $N\times M_A$ matrix $A$:
\begin{eqnarray}
\footnotesize{
A\equiv
\begin{pmatrix}
A_{1,1} & \cdots & A_{1,k} & \cdots & A_{1,M_A}\\
\vdots & \ddots &        &        & \vdots \\
A_{k,1} &        & A_{k,k} &        & A_{k,M_A} \\
\vdots &        &        & \ddots & \vdots \\
A_{N,1} & \cdots & A_{N,i} & \cdots & A_{N,M_A}
\end{pmatrix}}
\nonumber
\end{eqnarray}
and
$N\times M_B$ matrix $B$:
\begin{eqnarray}
\footnotesize{
B\equiv
\begin{pmatrix}
B_{1,1} & \cdots & B_{1,k} & \cdots & B_{1,M_B}\\
\vdots & \ddots &        &        & \vdots \\
B_{k,1} &        & B_{k,k} &        & B_{k,M_B} \\
\vdots &        &        & \ddots & \vdots \\
B_{N,1} & \cdots & B_{N,i} & \cdots & B_{N,M_B}
\end{pmatrix}}.
\nonumber
\end{eqnarray}
The matrix $(AB)$ is created by combining the columns of $A$ and $B$, defined as 
\begin{eqnarray}
\footnotesize{
(AB)\equiv
\begin{pmatrix}
A_{1,1} & \cdots  & A_{1,M_A} & B_{1,1} & \cdots &  B_{1,M_B} \\
\vdots &   \ddots       & \vdots &   \vdots    &    \ddots    & \vdots \\
A_{k,1}&   \cdots & A_{k,M_A}&  B_{k,1}  &        & B_{k,M_B}   \\
\vdots &    \ddots      & \vdots &     \vdots &  \ddots      &   \vdots \\
A_{N,1} & \cdots  & A_{N,M_A}&  B_{N,1} &  \cdots & B_{N,M_B}
\end{pmatrix}.}
\nonumber
\end{eqnarray}
Using $M_0$, we can calculate $S_{X_0}$ as $S_{X_0}=1-\mathrm{rank}_{[2]}(M_0)$
and find $S_{X_0\cup X_1}=2-\mathrm{rank}_{[2]}(M_0M_1)$ because the rank remains unchanged even if column swapping is performed.
Thus, we can similarly calculate $S_{X_0\cup X_1\cdots X_{j-1}}$ as $S_{X_0\cup X_1\cdots X_{j-1}}=j-\mathrm{rank}_{[2]}(M_0M_1\cdots M_{j-1})$.
To obtain the rank of $(M_1M_2\cdots M_{j-1})$,
we employ the Gaussian-elimination procedure and denote the result of 
performing the Gaussian-elimination procedure on $(M_1M_2\cdots M_{j-1})$ as $(\tilde{M_1}\tilde{M_2}\cdots \tilde{M_{j-1}})$, which is defined by over $F_2$.
Since the rank of submatrix $(\tilde{M_0}\cdots \tilde{M_k})$ is the same as the $\mathrm{rank}_{[2]} (M_0\cdots M_k)$, once we conduct the Gauss elimination to $(M_0M_1\cdots M_{L_{AB-1}})$, we can obtain $S_{X_0\cup X_1\cdots X_{j-1}}$ for all $j$.
We can apply the same discussion to the second term in Eq.~(\ref{s_contour}) 
and efficiently calculate $s_{AB}$.



\begin{thebibliography}{99}

\bibitem{Li2018}
Y. Li, X. Chen, and M. P. A. Fisher, Phys. Rev. B {\bf 98}, 205136 (2018).

\bibitem{Skinner2019}
B. Skinner, J. Ruhman, and A. Nahum, Phys. Rev. X {\bf 9}, 031009 (2019).

\bibitem{Li2019}
Y. Li, X. Chen, and M. P. A. Fisher, Phys. Rev. B {\bf 100}, 134306 (2019).

\bibitem{Vasseur2019}
R. Vasseur, A. C. Potter, Y. -Z. You, and A. W. W. Ludwig, Phys. Rev. B {\bf 100}, 134203 (2019).

\bibitem{Chan2019}
A. Chan, R. M. Nandkishore, M. Pretko, and G. Smith, Phys. Rev. B {\bf 99}, 224307 (2019).

\bibitem{Szyniszewski2019}
M. Szyniszewski, A. Romito, and H. Schomerus, Phys. Rev. B {\bf 100}, 064204 (2019).

\bibitem{Choi2020}
S. Choi, Y. Bao, X.-L. Qi, and E. Altman, Phys. Rev. Lett. {\bf 125}, 030505 (2020).

\bibitem{Bao2020}
Y. Bao, S. Choi, and E. Altman, Phys. Rev. B {\bf 101}, 104301 (2020).

\bibitem{Jian2020}
C. -M. Jian, Y.-Z. You, R. Vasseur, and A. W. Ludwig, Phys. Rev. B {\bf 101}, 104302 (2020).

\bibitem{Gullans2020}
M. J. Gullans, and D. A. Huse, Phys. Rev. X {\bf 10}, 041020 (2020).

\bibitem{Zabalo2020}
A. Zabalo, M. J. Gullans, J. H. Wilson, S. Gopalakrishnan, D. A. Huse, and J. H. Pixley, Phys. Rev. B {\bf 101}, 060301 (2020).

\bibitem{Sang2021}
S. Sang and T. H. Hsieh, Phys. Rev. Res. {\bf 3}, 023200 (2021).

\bibitem{Sang2021_v2}
S. Sang, Y. Li, T. Zhou, X. Chen, T. H. Hsieh, and M. P. A. Fisher, PRX Quantum {\bf 2}, 030313 (2021).

\bibitem{Nahum2021}
A. Nahum, S. Roy, B. Skinner, and J. Ruhman, PRX Quantum {\bf 2}, 010352 (2021).

\bibitem{Sharma2022}
S. Sharma, X. Turkeshi, R. Fazio, and M. Dalmonte, SciPost Phys. Core {\bf 5}, 023 (2022).

\bibitem{Fisher2022_rev}
M. P. A. Fisher, V. Khemani, A. Nahum, and S. Vijay, Annu. Rev. Condens. Matter Phys. 14:1, 335-379 (2023).


\bibitem{Block2022} 
M. Block, Y. Bao, S. Choi, E. Altman, and N. Y. Yao, Phys. Rev. Lett. {\bf 128}, 010604 (2022).

\bibitem{Richter2023}
J. Richter, O. Lunt, and A. Pal, Phys. Rev. Res. {\bf 5}, L012031 (2023).

\bibitem{Sierant2023}
P. Sierant, M. Schirò, M. Lewenstein, and X. Turkeshi, arXiv:2306.04764.

\bibitem{Fuji2020}
Y. Fuji, and Y. Ashida, Phys. Rev. B {\bf 102}, 054302 (2020).

\bibitem{Lunt2020}
O. Lunt and A. Pal, Phys. Rev. Res. {\bf 2}, 043072 (2020).

\bibitem{Goto2020}
S. Goto, and I. Danshita, Phys. Rev. A {\bf 102}, 033316 (2020).

\bibitem{Tang2020}
Q. Tang, and W. Zhu, Phys. Rev. Research {\bf 2}, 013022 (2020).

\bibitem{Turkeshi2021}
X. Turkeshi, A. Biella, R. Fazio, M. Dalmonte, and M. Schiró, Phys. Rev. B {\bf 103}, 224210 (2021).

\bibitem{Kells2022}
G. Kells, D. Meidan, and A. Romito, arXiv:2112.09787.

\bibitem{Fleckenstein2022}
C. Fleckenstein, A. Zorzato, D. Varjas, E. J. Bergholtz, J. H. Bardarson, and A. Tiwari, Phys. Rev. Res. {\bf 4}, L032026 (2022).

\bibitem{KOH2022}
Y. Kuno, T. Orito, and I. Ichinose, Phys. Rev. B. {\bf 106}, 214304 (2022).

\bibitem{Hayden2007}
P. Hayden and J. Preskill, J. High Energy Phys. {\bf 2007} (09), 120.

\bibitem{Hosur2016}
P. Hosur, X. L. Qi, D. A. Roberts, and B. Yoshida, J. High Energy Phys. {\bf 2016}, 1 (2016).

\bibitem{Xu2022}
S. Xu and B. Swingle, arXiv:2202.07060.

\bibitem{Gottesman1997}
D. Gottesman, arXiv:9807006.

\bibitem{Lang2020}
N. Lang, and H. P. Büchler, Phys. Rev. B {\bf 102}, 094204 (2020).

\bibitem{Ippoliti2021}
M. Ippoliti, M. J. Gullans, S. Gopalakrishnan, D. A. Huse, and V. Khemani, Phys. Rev. X {\bf 11}, 011030 (2021).


\bibitem{Zhu2023}
G. -Y. Zhu, N. Tantivasadakarn, S. Trebst, arXiv:2303.17627.

\bibitem{Lavasani2021}
A. Lavasani, Y. Alavirad, and M. Barkeshli, Nat. Phys. {\bf 17}, 342 (2021).

\bibitem{Klocke2022}
K. Klocke and M. Buchhold, Phys. Rev. B {\bf 106}, 104307 (2022).

\bibitem{KI2023}
Y. Kuno and I. Ichinose, Phys. Rev. B {\bf 107}, 224305 (2023).

\bibitem{Lavasani2021_2}
A. Lavasani, Y. Alavirad, and M. Barkeshli, Phys. Rev. Lett. {\bf 127}, 235701 (2021).

\bibitem{Negari2023}
A.-R. Negari, S. Sahu, and T.H. Hsieh, arXiv:2307.02292.

\bibitem{Sekino2008}
Y. Sekino and L. Susskind, Fast scramblers, J. High En-
ergy Phys. 2008 (10), 065

\bibitem{Maldacena2016}
J. Maldacena, S. H. Shenker, and D. Stanford, Journal of High Energy Physics {\bf 2016}, 106 (2016).

\bibitem{Bentsen2020}
G. Bentsen, T. Hashizume, A.S. Buyskikh, J. Davis, A.J. Daley, S.S. Gubser, and M. Schleier-smith, Phys. Rev. Lett. {\bf 123}, 130601 (2019).

\bibitem{Belyansky2020}
R. Belyansky, P. Bienias, Y.A. Kharkov, A. V. Gorshkov, and B. Swingle, Phys. Rev. Lett. {\bf 125}, 130601 (2020).

\bibitem{Yao2016}
N. Y. Yao, F. Grusdt, B. Swingle, M. D. Lukin, D. M. Stamper-Kurn, J. E. Moore, and E. A. Demler, arXiv:1607.01801.

\bibitem{Li2020}
Z. Li, S. Choudhury, and W. V. Liu, Phys. Rev. Res. {\bf 2}, 043399 (2020).

\bibitem{Li2022}
S. -S. Li, R. -Z. Huang, and H. Fan, Phys. Rev. B {\bf 106}, 024309 (2022).

\bibitem{Kuriyattil2023}
S. Kuriyattil, T. Hashizume, G. Bentsen, and A. J. Daley, arXiv:2304.09833.

\bibitem{Hashizume2022}
T. Hashizume, G. Bentsen, and A. J. Daley, Phys. Rev. Research {\bf 4}, 013174 (2022).

\bibitem{Hashizume2022_2}
T. Hashizume, S. Kuriyattil, A.J. Daley, and G. Bentsen, Symmetry (Basel). {\bf 14}, 1 (2022).

\bibitem{Sriram2022}
A. Sriram, T. Rakovszky, V. Khemani, and M. Ippoliti, arXiv:2207.07096.

\bibitem{Lucas2019}
A. Lucas, arXiv:1903.01468.

\bibitem{Aaronson2004}
S. Aaronson and D. Gottesman, Phys. Rev. A {\bf 70}, 052328 (2004).

\bibitem{Nielsen_Chuang}
M. A. Nielsen and I. L. Chuang, Quantum Computation and Quantum Information, by Michael A. Nielsen , Isaac L. Chuang, Cambridge, UK: Cambridge University Press, 2010 (2010).

\bibitem{Fattal2004}
D. Fattal, T. S. Cubitt, Y. Yamamoto, S. Bravyi, and I. L. Chuang, arXiv:0406168.

\bibitem{Nahum2017}
A. Nahum, J. Ruhman, S. Vijay, and J. Haah, Phys. Rev X {\bf 7}, 031016 (2017).


\bibitem{Shi2020}
B. Shi, X. Dai, and Y. -M. Lu, arXiv: 2012.00040.

\bibitem{Groisman2005}
B. Groisman, S. Popescu, and A. Winter, Phys. Rev. A {\bf 72}, 032317 (2005).

\bibitem{MacCormack2021}
I. MacCormack, M. T. Tan, J. Kudler-Flam, and S. Ryu, Phys. Rev. B {\bf 104}, 214202 (2021).

\bibitem{Chen2014}
Y. Chen and G. Vidal, Journal of Statistical Mechanics: Theory and Experiment 2014, {\bf 10011} (2014).

\bibitem{MacCormack2020}
J. Kudler-Flam, H. Shapourian, and S. Ryu, SciPost Physics {\bf 8}, 063 (2020). 


\bibitem{Sang2022}
S. Sang, Z. Li, T.H. Hsieh, and B. Yoshida, arXiv:2212.10634.


\bibitem{Bentsen2019}
G. Bentsen, Y. Gu, and A. Lucas, Proc. Natl. Acad. Sci. U. S. A. {\bf 116}, 6689 (2019).

\end{thebibliography}
\end{document}